# Experimental Investigation of Optimum Beam Size for FSO Uplink

Hemani Kaushal[a,*], Georges Kaddoum[b], Virander Kumar Jain[c], Subrat Kar[c]

[a] *Department of Electrical, Electronics and Communication Engineering, The NorthCap University, Gurgaon, Haryana, India.*
[b] *Département de génie électrique, École de technologie supérieure, Montréal (QC), Canada.*
[c] *Electrical Engineering Department, Indian Institute of Technology Delhi, New Delhi, India.*

**Abstract**

In this paper, the effect of transmitter beam size on the performance of free space optical (FSO) communication has been determined experimentally. Irradiance profile for varying turbulence strength is obtained using optical turbulence generating (OTG) chamber inside laboratory environment. Based on the results, an optimum beam size is investigated using the semi-analytical method. Moreover, the combined effects of atmospheric scintillation and beam wander induced pointing errors are considered in order to determine the optimum beam size that minimizes the bit error rate (BER) of the system for a fixed transmitter power and link length. The results show that the optimum beam size for FSO uplink depends upon Fried parameter and outer scale of the turbulence. Further, it is observed that the optimum beam size increases with the increase in zenith angle but has negligible effect with the increase in fade threshold level at low turbulence levels and has a marginal effect at high turbulence levels. Finally, the obtained outcome is useful for FSO system design and BER performance analysis.

## 1. Introduction

Laser beam propagation through random medium such as atmosphere or underwater has been studied extensively for many years [1–6]. This technology has several advantages over traditional radio frequency (RF) or microwave communication owing to its increased power efficiency, low mass and space requirement, higher directivity (i.e., larger antenna gain), and tariff-free extremely high bandwidth [7]. Also, FSO link provides high security due to beam confinement within a limited area and is less sensitive to electromagnetic interferences. However, various phenomena in the atmosphere such as absorption, scattering and turbulence limit the performance of FSO link. Out of these three effects, turbulence in the atmosphere can critically influence the quality of optical beam and degrade the system performance. In the case of ground-to-satellite optical communication system, the major concern is due to turbulence-induced beam wander and scintillation effects [8, 9]. The beam wander induced by the turbulence causes the beam to deflect from its original line-of-sight (LOS) path leading to pointing error displacement which further induces a change in the scintillation index. The latter degrades the BER performance of the system [10]. The combined effect of beam wander and intensity scintillation due to atmospheric turbulence on the BER performance of the system is significantly affected by the transmitted beam size and zenith angle. Titterton [11] was the first author to deal with scintillation and beam wander simultaneously and showed that the irradiance fluctuations with combined effect were substantially larger than caused by scintillation alone. The effect of the atmosphere on the optical beam is different for uplink and downlink paths. In case of FSO uplink from ground-to-satellite, the transmitter is in the vicinity of the atmosphere while the receiver is in the far-field. So for the uplink scenario, the transmitter beam size is often smaller than the outer scale of the turbulence $L_0$. This will cause the instantaneous point of maximum irradiance, known as "hot spot", to get displaced from its on-axis position. This effectively leads to pointing error displacement of the beam that causes the beam to miss the target. If the eddies size is smaller than the beam size, then a small portion of the beam will be diffracted and scattered independently. The movements of hot spot and short-term beam centroid will effectively lead to larger outer circle over a long period of time and is termed as long-term spot size $W_{LT}$ which is expressed as [12]:

$$W_{LT}^2 = W_{ST}^2 + \langle r_c^2 \rangle . \tag{1}$$

*Corresponding author
*Email address:* himaniz@yahoo.com (Hemani Kaushal)

In the above expression, $W_{ST}$ is the short term spot size which depends on two factors i.e., free space diffraction spreading and additional spreading caused by the turbulent eddies of size smaller than the beam size. The term $\langle r_c^2 \rangle$ describes the beam wander displacement variance caused by the large-size turbulent eddies. Therefore, it is seen that the $W_{LT}$ arises from the effects of turbulent eddies of all scale sizes. The use of Gaussian filter function can eliminate the small scale effect that leads to short term spot size such that the only contribution is due to beam wander effect. Therefore, $L_0$ forms an upper bound on the inhomogeneity size that results in beam wander [13]. It may be mentioned that this effect is negligible for FSO downlink from the satellite as when the beam reaches the atmosphere, its size is much larger than the turbulent eddy size and that would not displace the beam centroid significantly.

Most of the previous literature have carried out a theoretical analysis of combined effect of atmospheric scintillation and pointing errors due to beam wander effect. The BER performance of ground-to-satellite FSO uplink using beam wander effect has been studied in [14]. In [15], the authors have presented a theoretical study of scintillation and beam wander analysis for ground-to-satellite uplink communication using fast-Fourier-transform beam propagation method. The spatial and temporal statistics of beam wander effect for FSO uplink have been studied analytically in [16]. In [17] and [18], the authors have compared the theoretical and simulated results for analyzing the increase in scintillation due to the effect of beam wander for 10 km horizontal path propagation at high altitude. Another simulation based study on the effect of beam wander on the BER performance of FSO system using collimated and focused Gaussian beams have been carried out in [19]. A theoretical study of the influence of beam wander on ground-to-satellite FSO uplink for a collimated untracked Gaussian beam and its impact on BER system performance is given in [20]. The experimental study on beam wander effect under varying atmospheric turbulence was carried out in [8]. Later in [21], the effect of spatial coherence of a Gaussian Schell-model (GSM) beam on beam wander was studied experimentally and it was demonstrated that GSM beam with low coherence experience small beam wander effect. In [22], optical spatial filter was proposed to reduce the effect beam wander caused due to turbulence in the atmosphere.

To the best of authors ' knowledge, the experimental investigation of optimum transmitter beam size for FSO uplink considering beam wander effect is carried out first time in this paper. Based on gamma-gamma distribution model of irradiance fluctuations, probability density function (pdf) of received irradiance is computed using a semi-analytical method by varying the temperature and wind speed inside the OTG chamber. The experimental results are then used to determine the optimum beam size that minimizes the BER performance of the system. Our work will be beneficial for meeting the design requirements in FSO uplink communication system.

The rest of the paper is organized as follows: Section 2 describes the channel model and the effect of beam wander on ground-to-satellite FSO uplink. Experimental description and characterization of atmospheric turbulence are discussed in Section 3. Finally, the experimental results and conclusions are presented in Sections 4 and 5, respectively.

## 2. Atmospheric Turbulence and Beam Wander Effect

For ground-to-satellite FSO communication, the laser beam has to pass through random inhomogeneities in the atmosphere arising due to refractive index fluctuations. These fluctuations are described by a parameter called 'refractive index structure parameter' $C_n^2(h)$ which gives the strength of turbulence in the atmosphere as a function of height above the ground. It is the most critical parameter that can be obtained by measuring temperature ($T$), pressure ($P$), wind speed ($V$) and temperature spatial fluctuations ($\Delta T$) along the propagation path and is given as [12]:

$$C_n^2 = \left[79 \times 10^{-6} \frac{P}{T^2}\right]^2 C_T^2, \tag{2}$$

where $C_T^2$ is the temperature structure parameter which is determined by taking the measurements of mean square temperature between two points separated by a certain distance along the propagation path (in $\mathrm{deg}^2/\mathrm{m}^{2/3}$) and is expressed as:

$$C_T^2 = \left\langle \Delta T^2 \right\rangle r^{-1/3}, \tag{3}$$

where $\Delta T = T_1 - T_2$ ($T_1$ and $T_2$ are the temperatures of two arbitrary point separated by a distance $r$) and the angle bracket $\langle\rangle$ denotes an ensemble average. The profile model of $C_n^2$ describes the variation in structure parameter as a function of altitude $h$. Several day time and night time profile models are used for FSO communication [7]. One of the most popular empirical model is Hufnagel-Valley Boundary (HVB) model described by:

$$\begin{aligned} C_n^2(h) \;=\; & 0.00594 \left[ \left(\frac{V}{27}\right)^2 \left(10^{-5} h\right)^{10} \exp\left(-h/1000\right) \right. \\ & +2.7 \times 10^{-16} \exp\left(-\frac{\mathrm{h}}{1500}\right) \\ & \left. +A \exp\left(-\frac{h}{100}\right)\right] m^{-2/3}, \end{aligned} \tag{4}$$

where $V$ is the RMS wind speed in m/s and $A$ is the value of refractive index structure parameter at the ground i.e., $C_n^2(0)$. The presence of turbulence in the atmosphere will lead to the formation of eddies of different refractive indices and various sizes. Depending on the size of turbulent eddies and transmitter beam size, three types of atmospheric turbulence effects can be identified: beam wander, beam scintillation and beam spreading. In FSO uplink, a beam with size smaller than $L_0$ will be deflected as a whole in random manner from its original path. This leads to significant change in the direction of the beam resulting in beam wander effect which is characterized statistically by the variance of beam wander displacement $\langle r_c^2 \rangle$ and expressed as [17]:

$$\langle r_c^2 \rangle = 0.54\,(H - h_0)^2 \sec^2(\theta) \left(\frac{\lambda}{2W_0}\right)^2 \left(\frac{2W_0}{r_0}\right)^{5/3}, \tag{5}$$

where $\theta$ is the zenith angle, $\lambda$ is the operating wavelength, $W_0$ is the transmitter beam radius, $H$ and $h_0$ are the altitude of satellite and transmitter, respectively. For ground based transmitter, $h_0 = 0$ and satellite altitude $H = h_0 + L\cos(\theta)$, where $L$ is propagation length. The other parameter $r_0$ is the atmospheric coherence length (or Fried parameter) defined by:

$$r_0 = \left[0.423k^2 \sec(\theta) \int_{h_0}^{H} C_n^2(h)\,dh\right]^{-3/5}, \tag{6}$$

where $k = 2\pi/\lambda$ is the optical wave number. The beam wander effect will lead to effective pointing error of the beam $\sigma_{pe}$ given as [17]:

$$\sigma_{pe}^2 = \langle r_c^2 \rangle \left[1 - \left(\frac{C_r^2 W_0^2/r_0^2}{1 + C_r^2 W_0^2/r_0^2}\right)^{1/6}\right]. \tag{7}$$

In the above equation, the parameter $C_r$ is a scaling constant typically in the range from 1 to $2\pi$. Fig.1 shows the FSO uplink propagation model for Gaussian beam illustrating $W_0$, spot size $W(L)$, mean transmit intensity $I(r, 0)$, mean received intensity $I(r, L)$ and angular pointing error $\alpha_r$. Beam wander induced pointing error will further increase the scintillation index whose value will be different from that predicted by conventional Rytov theory [20]. According to the first-order Rytov theory, the scintillation index $\sigma_I^2$ is expressed as sum of longitudinal $\sigma_{I,l}^2(L)$ and radial $\sigma_{I,r}^2(r, L)$ components. However, when considering beam wander effect for FSO uplink, the increase in $\sigma_{I,l}^2(L)$ should be considered as the effect of beam wander may increase the long-term beam profile near the bore-sight and result in a slightly flattened beam. Therefore, for FSO uplink, the longitudinal scintillation index consists of two components i.e., scintillation produced by beam wander effect and scintillation produced due to longitudinal displacement (i.e., due to intensity fluctuations of the beam not subject to beam wander effect) which is expressed as [17]:

$$\sigma_{I,l}^2(L) = 5.95\,(H-h_0)^2 \sec^2(\theta) \left(\frac{2W_0}{r_0}\right)^{5/3} \left(\frac{\alpha_{pe}}{W}\right)^2 + \sigma_{R'}^2. \tag{8}$$

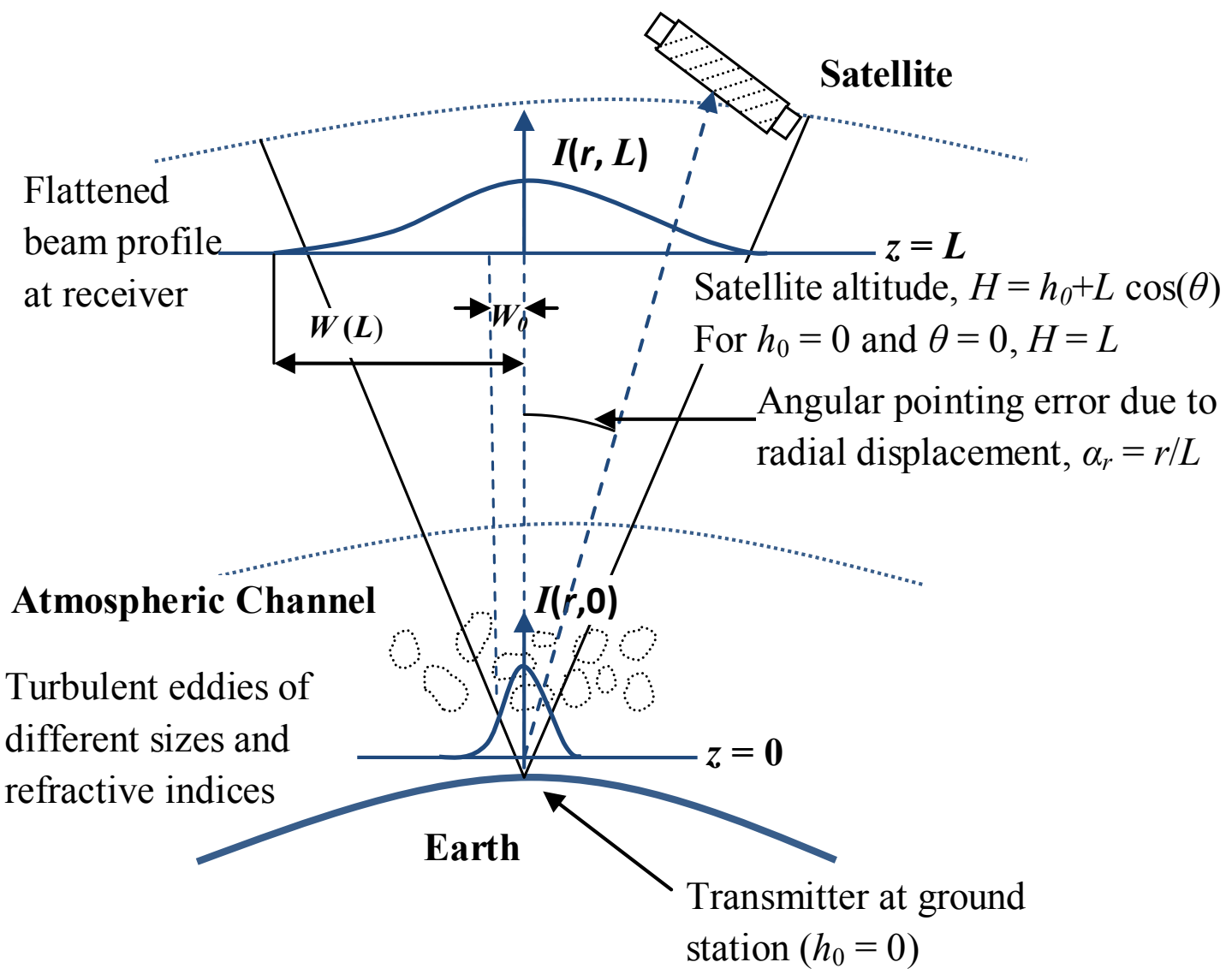

Figure 1: Beam transmission propagation model for ground-to-satellite FSO uplink.

In the above equation, $\alpha_{pe} = \sigma_{pe}/L$ is the angular pointing error due to beam wander and $\sigma_{R'}^2$ is the Rytov variance of the Gaussian beam without considering beam wander effect which is given as [17]:

$$\sigma_{R'}^2 = 8.70\mu_1 k^{7/6} (H-h_0)^{5/6} \sec^{11/6}(\theta), \tag{9}$$

where

$$\begin{aligned} \mu_1 &= \mathrm{Re}\int_{h_0}^{H} C_n^2(h) \Big\{ (\xi)^{5/6} \left[\Lambda\xi + i\,(1 + L\xi/F)\right]^{5/6} \\ &\quad - \Lambda^{5/6}\xi^{5/3} \Big\}\, dh. \end{aligned} \tag{10}$$

Here, $\Lambda = 2L/kW^2$, $\xi = 1-(h-h_0)/(H-h_0)$, $W$ and $F$ are the beam spot size and radius of curvature, respectively as viewed in the receiver plane. Similarly, the radial component of scintillation is given as [17]:

$$\begin{aligned} \sigma_{I,r}^2(r,L) &= 5.95\,(H-h_0)^2 \sec^2(\theta) \left(\frac{2W_0}{r_0}\right)^{5/3} \\ &\quad \times \left(\frac{\alpha_r - \alpha_{pe}}{W}\right) U(\alpha_r - \alpha_{pe}) + \sigma_{I,l}^2(L), \end{aligned} \tag{11}$$

where $\alpha_r = r/L$ and $U(x)$ is a unit step function. For small radial displacement i.e., $r < \sigma_{pe}$, the scintillation index $\sigma_{I,r}^2(r,L) = \sigma_{I,l}^2(L)$. These expressions are later used in Section 3 to study the variation of scintillation index as a function of transmitter beam size for various turbulence strengths generated inside OTG chamber. Also, in order to ensure that the detected fluctuations at the receiver is due to wander-induced pointing error displacement, the beam diameter, $D$ ($=2W_0$) should be less than the atmospheric coherence length ($r_0$). For $D \gg r_0$, the wandering effect will no longer exist as the beam will break up into smaller pieces due to loss of spatial coherence across the beam.

## 3. Experimental Set-up and Turbulence Characterization

This section gives the description of the experimental setup used to generate turbulence inside laboratory environment. The setup is built in such a way that it allows the laser beam to propagate through closed turbulent chamber where the turbulence can be controlled by varying temperature or wind speed inside the chamber. Further, in order to emulate the beam propagation through decreasing temperature profile with the increase in altitude, the laser beam is allowed to pass through a vertical column where liquid nitrogen is poured from the top to build up a negative temperature gradient inside the column.

In order to create a real turbulence inside the OTG chamber, the framework has to fulfill some important scaling laws to ensure compatibility with the real atmospheric scenario. The atmosphere is modeled using Kolmogorov theory [12] which defines the outer scale ($L_0$) and inner scale ($l_0$) of the turbulent eddies. Therefore, in order to ensure realistic simulations, the same behavior has to be reproduced by the OTG chamber. For this, we need to work on dimensionless variables abiding certain scaling laws so that they fulfill the specifications that are applicable for FSO uplink from ground-to-satellite. The first scaling law is to maintain a constant Strehl ratio which is an important parameter for an optical imaging system. It is defined as the ratio of on-axis mean irradiance in turbulence of a point source to that in the absence of turbulence. For FSO uplink, we used the principle of reciprocity so that the far-field Strehl ratio is defined by near-field phase variance and is expressed as:

$$SR \approx \left[1 + \left(\frac{D_0}{r_0}\right)^{5/3}\right]^{-1}, \tag{12}$$

where $D_0^2 = 8W_0^2$ and $r_0$ is the Fried parameter. Preserving this ratio will make it possible to simulate the behavior of large telescope in real scenario with a small one in laboratory environment. For achieving this ratio, we considered a scenario of laser uplink communication from ground-to-satellite in NASA's lunar laser communication demonstration (LLCD) program [23]. Lunar lasercom ground system consists of an array of four 15 cm uplink telescopes operating between 1550 nm and 1579 nm. Using Eq. (12) for 15 cm transmitter telescope at 1550 nm with $r_0$= 19 cm (typical value at 1550 nm), Strehl ratio comes out to be 0.45. Maintaining this ratio for transmitter diameter $D$ of 6 mm, $r_0$ of about 7.6 mm is produced by the OTG chamber. Similarly, there is need to match the outer scale and inner scale generated inside the OTG chamber for the same magnitude as that in the atmosphere. In order to achieve that, the scaling law for inertial sub-range (family of eddies bounded above by $L_0$ and below by $l_0$) i.e., $D/L_0$ and $D/l_0$ have to be constant. For a 15 cm uplink telescope, the ratio of $D/L_0$ ranges from 0.025 to 0.15 and $D/l_0$ from 15 to 150. A beam diameter of 6 mm when passed through OTG chamber results in the same ratio if $L_0$ varies from 240 mm to 40 mm and $l_0$ from 0.4 mm to 0.04 mm. The effect of inner scale damping will be negligible until care is taken to keep the ratio of $D/l_0$ at high values by choosing large beam diameters. An experimental setup for FSO communication consisting of a transmitter, atmospheric channel, and a receiver is shown in Fig. 2.

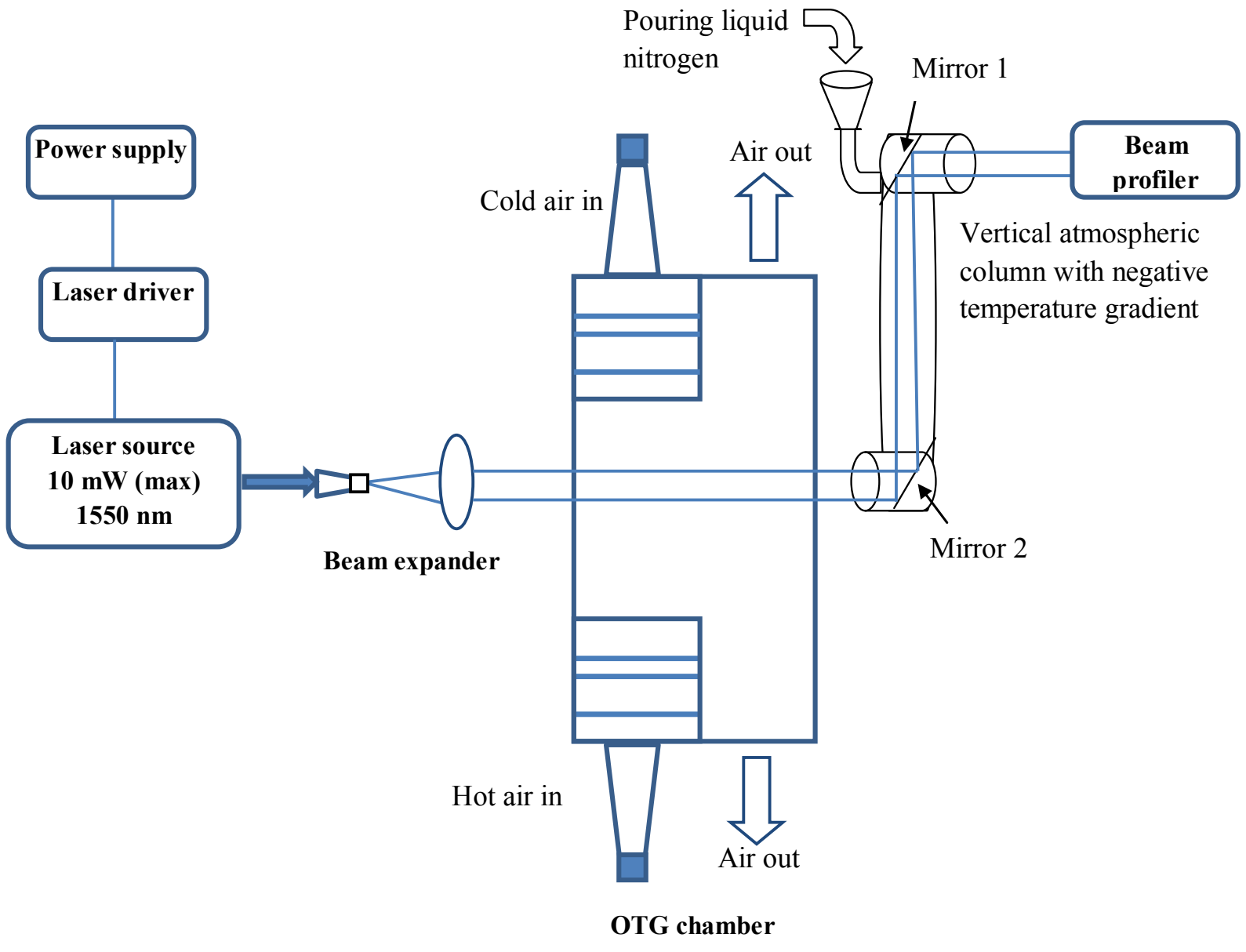

Figure 2: Experimental setup for measurement.

The transmitter used in the experiment is a laser source with maximum output power of 10 mW operating at a wavelength of 1550 nm. A beam expander is used to expand the width of transmitted beam from the laser source. This expanded beam is then passed through an atmospheric channel consisting of an OTG chamber which generates an artificial turbulence inside the laboratory environment by forced mixing of cold and hot air [8, 10, 24]. Honeycomb filters are used to allow laminar airflow inside the mixing zone of OTG chamber. A three dimensional and top view of OTG chamber is shown in Fig. 3(a) and (b), respectively.

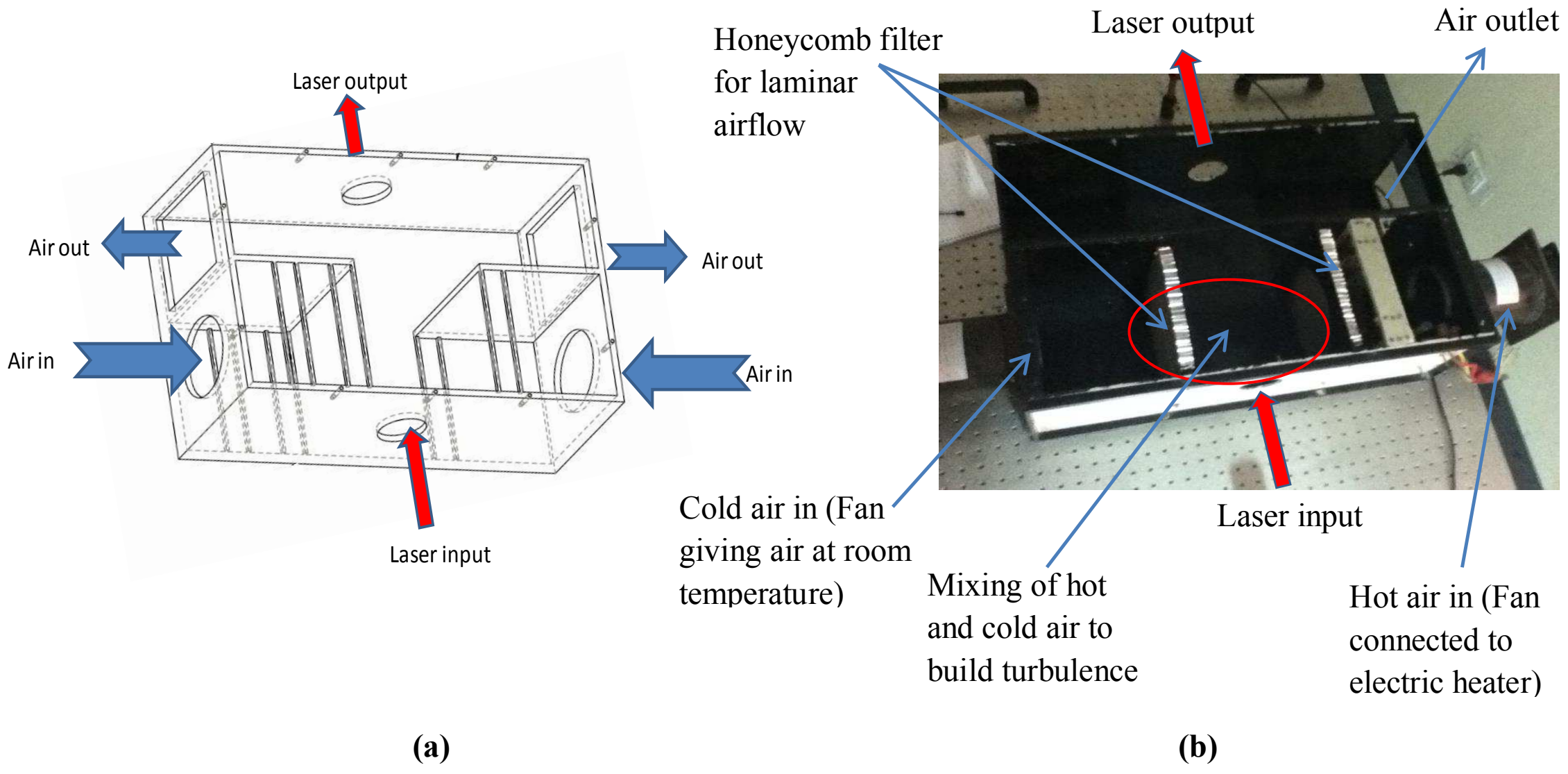

Figure 3: OTG chamber (a) Three dimensional view and (b) top view.

The dimensions of OTG mixing chamber are 20 x 20 x 20 cm which is compatible with determined outer scale of turbulence. However, the propagation length of laser beam inside the OTG chamber is increased by using a highly reflective mirror structure. Depending upon the placement of mirrors, the laser beam will travel back and forth through multiple reflections from the mirrors till it escapes from the chamber as shown in Fig. (4) (a) . Using this assembly, the propagation length is increased up to 5 m. The distance calibration is carried out at the beginning of the experiment using red laser (633 nm) by visual counting of multiple reflections inside the mirror assembly as well as by measuring the radius of the beam that leaves the assembly with the help of beam profiler that is connected to the computer for Matlab computation. The average radius of the beam is calculated by measuring the distance of $1/e^2$ times the maximum intensity point from the beam centroid. In this experiment, a 3 mm transmitter beam becomes 4.2 mm after multiple reflections up to 25 times inside the OTG chamber. This value is close enough to the theoretical value which is approx. 5.8 mm. Fig. (4) (b) shows the variation of effective beam spot size as a function of transmitter beam size after propagating a link length of 5 m inside the OTG chamber.

Different values of atmospheric turbulence can be generated by varying pressure, speed or temperature of the air flow inside the chamber. For this reason, the temperature inside the chamber is controlled by varying the variac connected to the heater. The air intake at one end of the channel is kept at room temperature while hot air is blown from the other side whose temperature is controlled using the electrical heater. The velocity inflow of the air is controlled by varying the rpm of the fans connected to the OTG chamber. By varying the variac connected to heater and rpm of fans, we are able to generate varying strength of turbulence inside the chamber. Laser beam propagation is kept perpendicular to the direction of the turbulent air flow. The aberrated wavefront after OTG chamber is passed through a vertical column which has a mirror arrangement inside it to allow vertical propagation of the optical beam. In order to create a negative temperature gradient as present in the atmospheric channel, we poured liquid nitrogen from the top of the vertical column. The optical beam from the vertical column is then captured by the beam profiler which has in-built high resolution CCD (charged coupled device) camera. The digitized images captured by the camera are transmitted to the computer via USB interface for further Matlab computations. The beam profiler not only provides basic beam information like beam width, relative beam power or energy distribution, peak pixel location, beam ellipticity, statistical measurements etc., but it also provides automatic image control features (i.e., automatic gain control (AGC), automatic frame control (AFC) and automatic exposure shutter (AES) which helps to reduce the noise in the camera and improve the quality of the captured image. Neutral density filter of transmissivity 0.3% is used before the beam profiler to avoid its saturation. The whole setup was fixed on a vibration-free table and the observations were taken in a dark room environment. The complete set of parameters used in the experiment is given in Table 1.

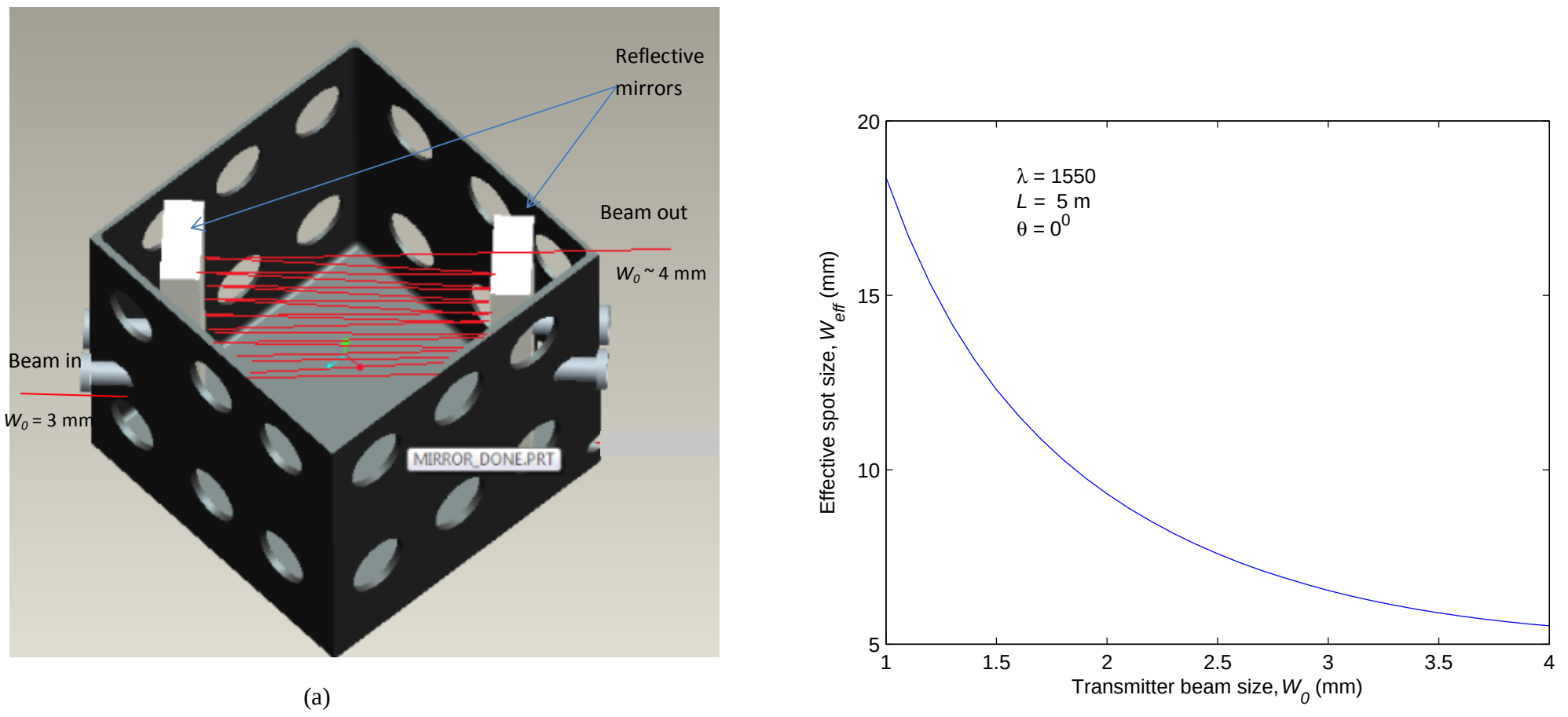

Figure 4: Mirror assembly: (a) structure placed inside the OTG chamber for multiple reflections (b) effective beam radius after leaving mirror assembly vs transmitter beam radius.

### *3.1. Turbulence Characterization*

In order to characterize the turbulence generated inside OTG chamber, experiments were carried out over several temperature ranges by varying the variac connected to the electrical heater. The wind speed was kept fixed at 3.5 m/s. To study the effect of turbulence-induced beam wander, fluctuations in the beam centroid were measured, firstly without OTG chamber turned on (i.e., no turbulence) and secondly with OTG chamber turned on at a fixed temperature and wind speed (i.e., with turbulence). Initially 150 shots are captured at room temperature and the centroid is calculated for each of these shots by taking the weighted average of the pixels with intensity as the weight. The fluctuations in the centroid are determined for these 150 shots before OTG chamber is turned on. The variance of these fluctuations is basically the background noise. This background noise is subtracted from the measured variance after

Table 1: Parameters used in laboratory experiment.

| Parameters | Values |
|---|---|
| Laser power | 10 mW |
| Operating wavelength ($\lambda$) | 1550 nm |
| Variable beam expander | 1X - 5X continuous expander |
| OTG chamber dimensions | 20 x 20 x 20 cm |
| Temperature difference ($\Delta T$) | 10 K to 90 K |
| Zenith angle | 0 degree |
| Propagation length ($L$) | 5 m |
| Wind speed ($V$) | 3.5 m/s |
| Receiver camera type | Beam profiler (CCD type) |
| Pixel size | 4.4 x 4.4 microns |
| Frame rate | 2 frames per sec |
| Exposure time | 0.4 ms |
| Camera resolution | 1280 x 1024 pixels |

OTG chamber is turned on so that the results obtained are due to turbulence only. The experiment was repeated for six different temperature differences and their corresponding values of variance were observed using beam profiler. In order to understand the effect of beam propagation through OTG chamber and vertical column with negative temperature gradient, the measurement of beam variance was carried out in two phases i.e, with and without creating negative temperature gradient profile inside the vertical column. It was observed that for the both cases the variance of beam wander displacement varies linearly with the temperature difference. However, the value of beam variance is more for the set up without having negative temperature gradient inside the vertical column. This is because the link without negative temperature gradient does not show much variation in $C_n^2$ as a function of height and is therefore, almost similar to horizontal link. On the other hand, the negative temperature gradient builds up a temperature difference within the vertical column so that the average value of $C_n^2$ when integrated over the entire path length decreases and thereby, reducing the value of beam wander variance. Fig. 5 shows the variation of beam variance with temperature differences for both the phases.

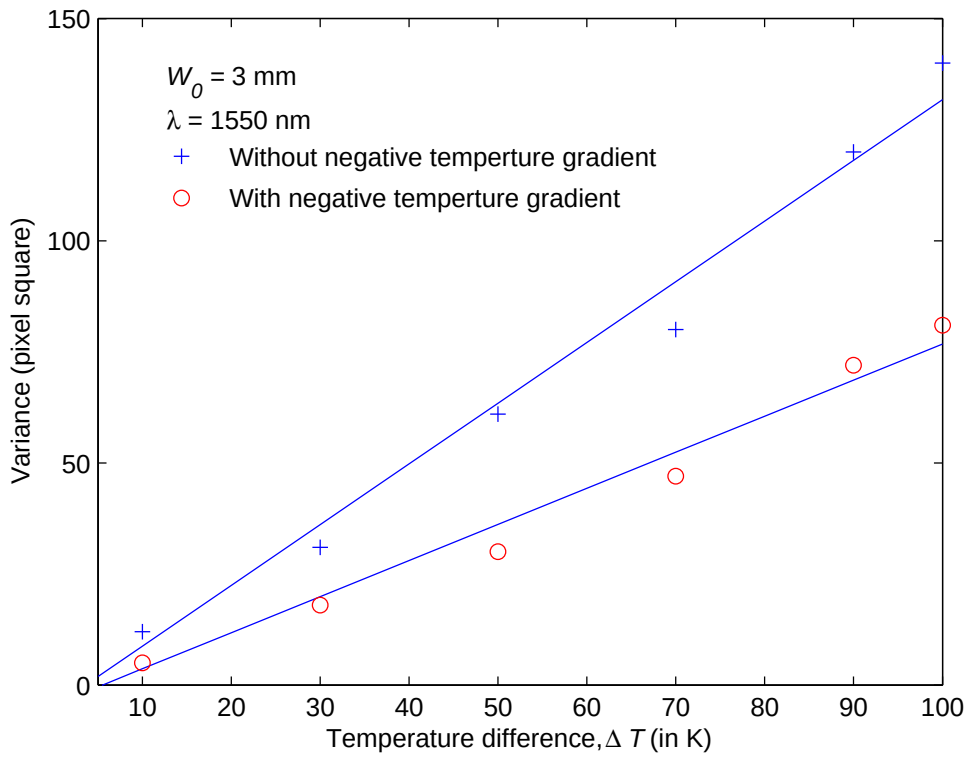

Figure 5: Beam wander variance vs temperature with and without negative temperature gradient profile inside vertical atmospheric column.

Once the beam wander variance is known, the experimental values of $C_n^2 \cdot H$ are determined using

Eqs. (5) and (6) for various temperature differences inside OTG chamber. The value of $C_n^2 \cdot H$ without OTG chamber turned on come out to be 1.6 x $10^{-15}$ $m^{1/3}$. The results of $C_n^2 \cdot H$ for other temperature differences when OTG chamber is turned on are shown in Table 2. It is seen from this table that the achieved scintillation index for small values of $\triangle T$ describes a weak turbulence condition despite of high values of $C_n^2 \cdot H$. This difference is attribute to the relatively short propagation path of 5 m inside the laboratory environment.

Table 2: $C_n^2 \cdot H$ values for $W_0$= 3 mm at wind speed of 3.5 m/s for various temperature differences.

| S. No. | Temperature difference $\Delta T$ (in K) | $C_n^2 \cdot H$ $\left(m^{1/3}\right)$ | Scintillation index $\sigma_{I,}^2(L)$ |
|---|---|---|---|
| 1. | 10 | $5.50\times10^{-13}$ | 0.0002 |
| 2. | 30 | $2.65\times10^{-12}$ | 0.0014 |
| 3. | 50 | $5.92\times10^{-12}$ | 0.0083 |
| 4. | 70 | $8.46\times10^{-12}$ | 0.0155 |
| 5. | 90 | $5.22\times10^{-11}$ | 0.79 |
| 6. | 100 | $6.72\times10^{-11}$ | 1.25 |

## 4. Results

Following the procedure described in Section 3, the variance of beam centroid due to beam wander effect and its corresponding $C_n^2 \cdot H$ values are observed for various temperature differences. After characterizing the turbulence inside OTG chamber, scintillation index is determined using Eqs. 7 to 11 for a given operating wavelength, $\lambda$ = 1550 nm, zenith angle, $\theta = 0^0$, transmitter height, $h_0 = 0$, transmitter beam size, $W_0$= 3 mm and propagation distance, $L$ = 5 m. These values are then used to plot PDF of the received irradiance using semi-analytical method for gamma-gamma distribution model which is given as [3]:

$$f_I(i) = \frac{2(\alpha\beta)^{\frac{\alpha+\beta}{2}}}{\Gamma(\alpha)\Gamma(\beta)\, i}\left(\frac{i}{\langle I(r,L)\rangle}\right)^{\frac{\alpha+\beta}{2}} \times K_{\alpha-\beta}\left(2\sqrt{\frac{\alpha\beta i}{\langle I(r,L)\rangle}}\right) \quad \text{for } (i>0), \tag{13}$$

where $\Gamma$ and $K$ are the Gamma function and modified Bessel function of the second kind, respectively. The parameters $\alpha$ and $\beta$ are expressed as:

$$\alpha = \left[\exp\left\{\frac{0.49\sigma_{R'}^2}{\left[1+(2+L/F)\,0.56\sigma_{R'}^2\right]^{7/6}}\right\}-1\right]^{-1} \tag{14}$$

and

$$\beta = \left[\exp\left\{\frac{0.51\sigma_{R'}^2}{\left[1+0.69\sigma_{R'}^{12/5}\right]^{7/6}}\right\}-1\right]^{-1}. \tag{15}$$

$\langle I(r,L)\rangle$ is the mean intensity in the presence of turbulence and is given by [12]:

$$\langle I(r,L)\rangle = \frac{W_0^2}{W_{LT}^2}\exp\left(-\frac{2r^2}{W_{LT}^2}\right), \tag{16}$$

where $W_{LT}$ is long term spot size which is formed by superposition of the instantaneous spots that reach the receiver and under weak fluctuations is given by [12]:

$$W_{LT} = W\sqrt{1+\left(8W_0^2/r_0^2\right)^{5/6}},\ 0 \leq \left(8W_0^2/r_0^2\right) \leq 1. \tag{17}$$

The pdf for received irradiance is shown in Fig. 6 for various values of temperature differences inside OTG chamber. It is observed that for a fixed transmitter beam size, increase in temperature difference (or turbulence) causes the pdf to shift towards the left till it becomes negative exponential at a higher temperature difference.

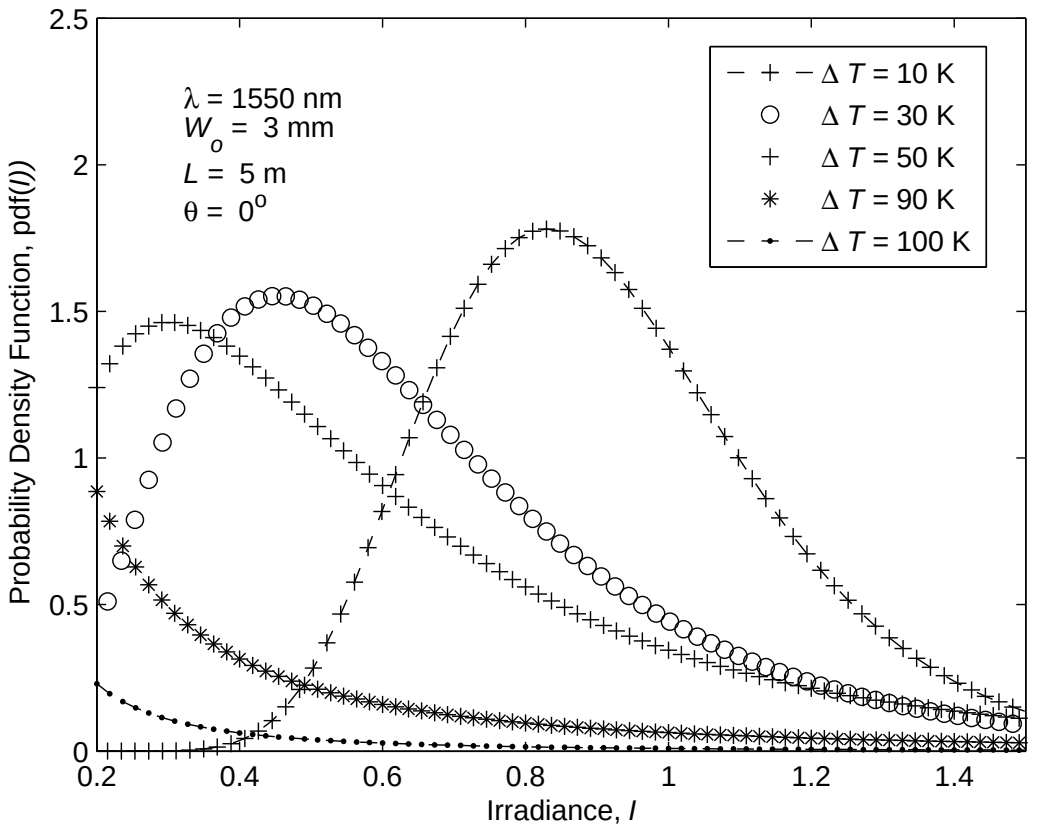

Figure 6: The pdf of received irradiance for a fixed transmitter beam size $W_0 = 3$ mm at various temperature differences.

Further, the pdf is observed at a fixed temperature of 10 K for different values of transmitter beam size. It was observed that the intensity distribution decreases with the increase in the transmitter beam size as shown in Fig. 7. A similar trend was seen at higher values of temperature differences except that in this case, the curve shifts towards the left till the distribution becomes negative exponential. This clearly indicates the significance of transmitter beam size on the statistical characteristics of the received signal.

Fig. 8 shows the effect of transmitter beam size on the longitudinal scintillation index $\sigma_{I,l}^2(L)$ for different values of temperature differences inside OTG chamber. It clearly indicates that there exists a particular beam size (between 2.5 mm to 4 mm) for which the beam variance due to turbulence-induced beam wander effect is minimum for given operating wavelength $\lambda$ and propagation length $L$. Also, a very slight difference is observed between a theoretical and experimental results for a link length of 5 m operating at 1550 nm and zero zenith angle. It is to be mentioned that the trend of curve will remain same at other operating wavelengths as well, however, the dip in scintillation index may be observed at some other value of transmitter beam size. Also, in case of long link distances that cannot be emulated in laboratory controlled environment e.g., ground-to-satellite uplink communication, large scale fading

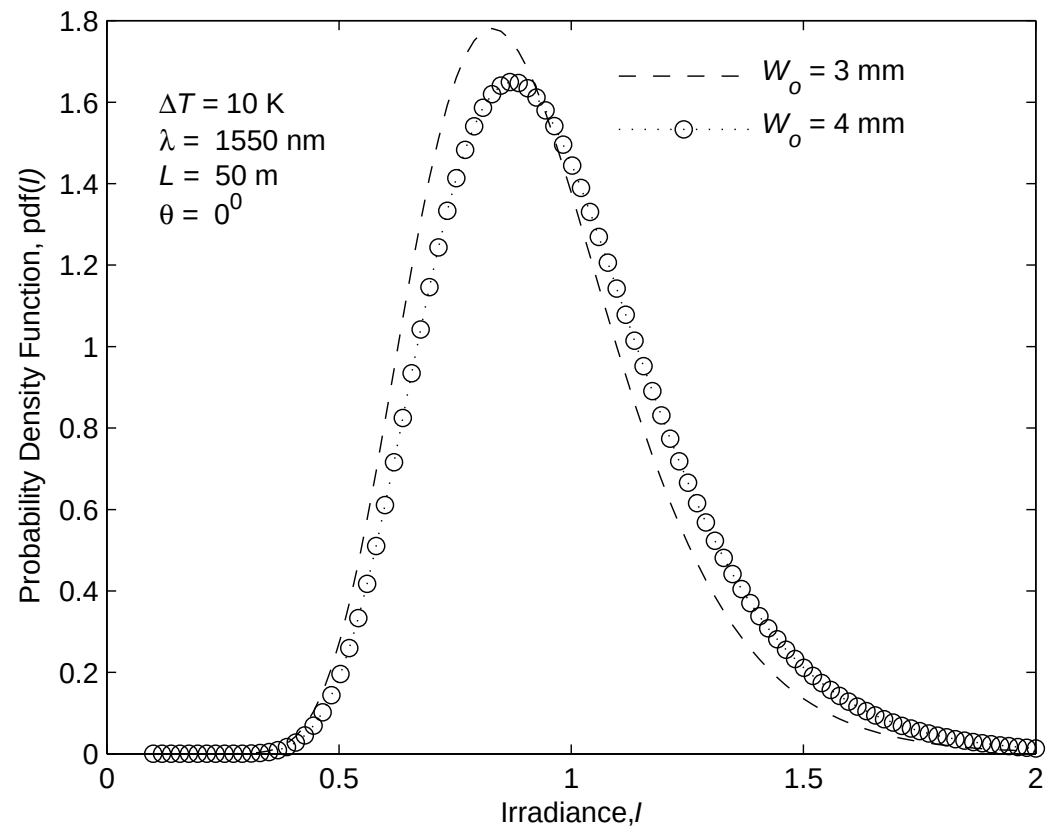

Figure 7: Variation in pdf of received irradiance with various transmitter beam sizes for a fixed propagation length $L = 5$ m and temperature difference $\Delta T = 10$ K.

effect caused due to free-space and beam divergence losses have to be considered.

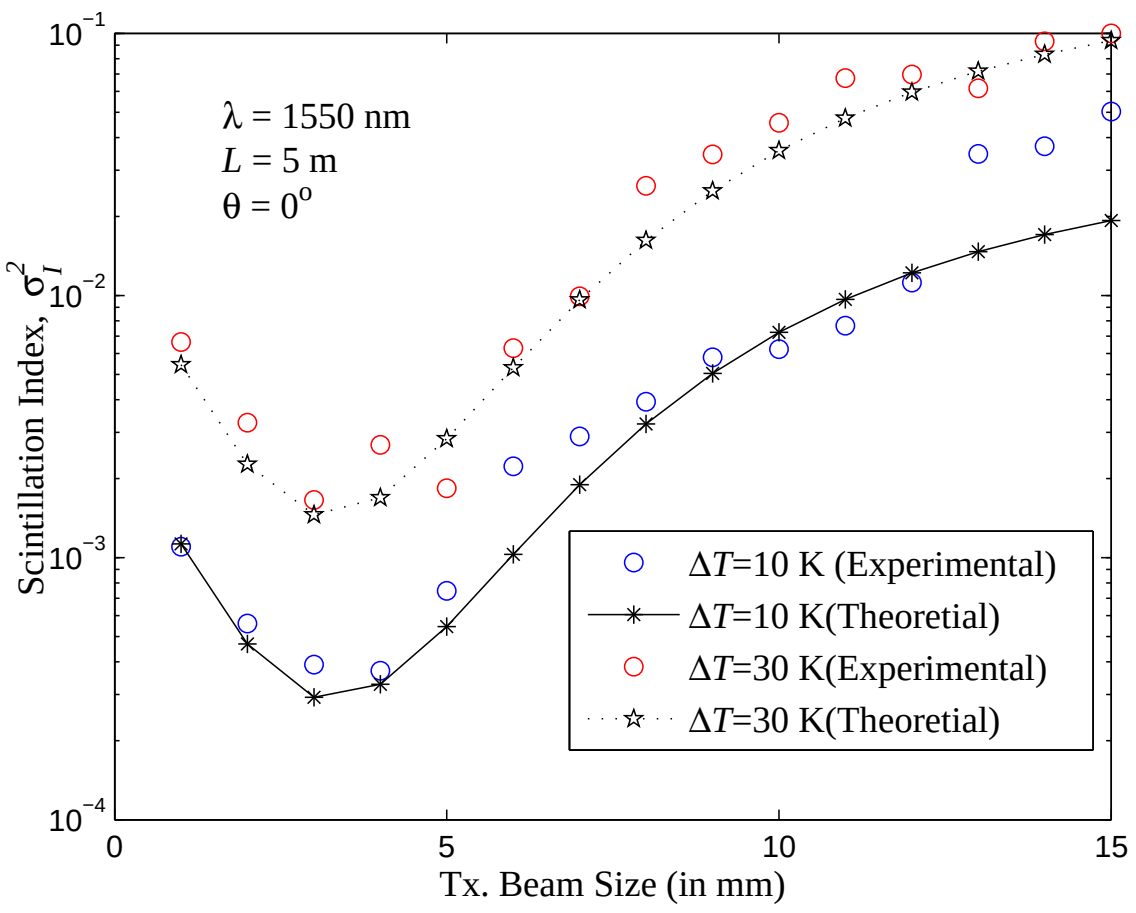

Figure 8: Longitudinal scintillation index as a function of transmitter beam size for different values of temperature differences inside OTG chamber.

Similarly, Fig. 9 shows the effect of radial displacement on the value of scintillation index for a given transmitter beam size. It is observed that there is very little effect on the scintillation index for a small propagation distance of 5 m. However, we see an increase in the value of scintillation index with the increase in turbulence level.

Since BER is a function of beam size and scintillation index, it is evident that there exists an optimal beam size that minimizes the BER of the system. For smaller beam size, beam wander induced pointing error increases, while for larger beam size, the value of scintillation index increases. Therefore, the combined effect of beam wander induced pointing error and scintillation index will help in achieving an optimum beam size that minimizes the BER of the system. For intensity modulated system with threshold $I_{Th}$, BER in the presence of turbulence is obtained by applying conditional averaging over

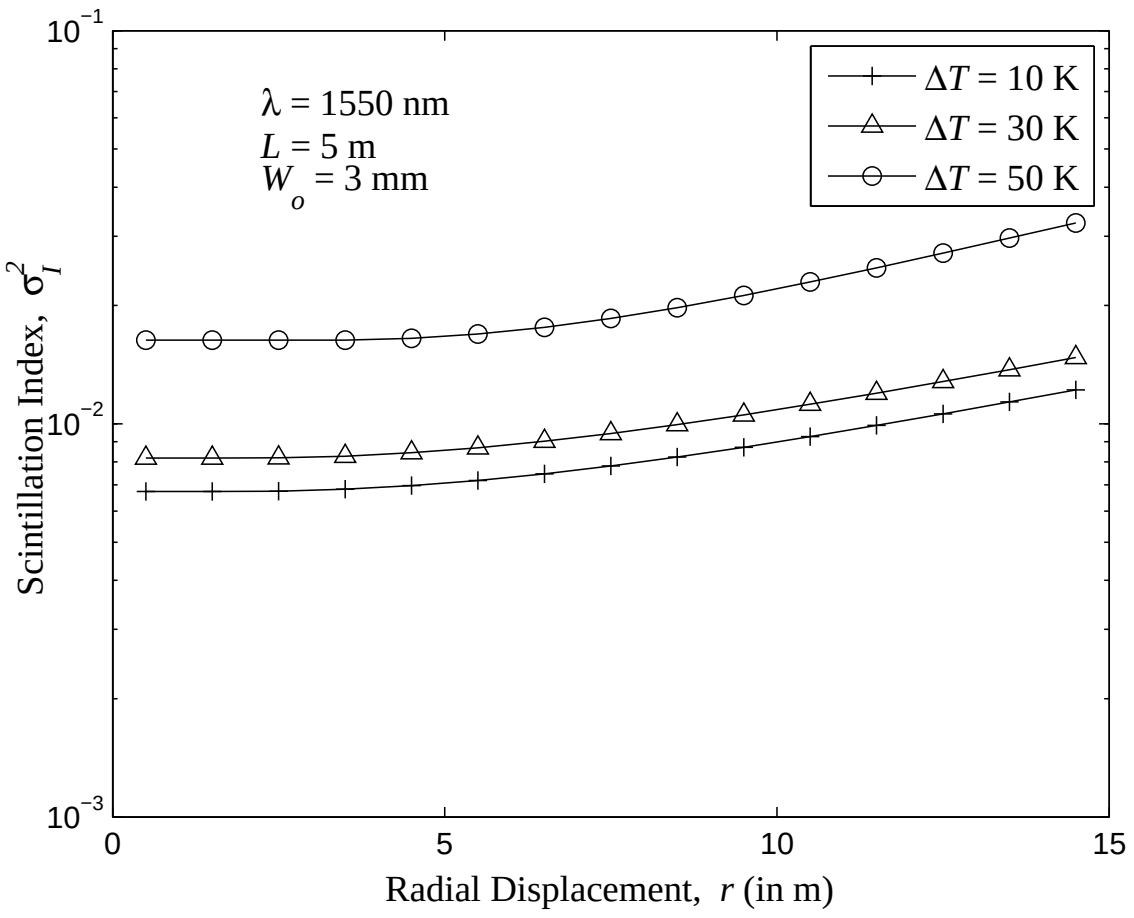

Figure 9: Scintillation index as a function of radial displacement from on-axis position for a beam size of 3 mm.

the fading parameter and is given by:

$$\mathrm{BER} = \frac{1}{2} P\left(I \leq I_{Th}\right) = \frac{1}{2} \int_0^{I_{Th}} f_I\left(I\right) dI. \tag{18}$$

Usually, fade threshold parameter $F_{Th}$ is used instead of intensity threshold which is given as:

$$F_{Th} = 10\log_{10}\left(\left\langle I\left(0,L\right)\right\rangle / I_{Th}\right). \tag{19}$$

Fig. 10 shows the variation of BER with transmitter beam size for different values of temperature differences inside the OTG chamber using semi-analytical method. This figure clearly shows a dip in the value of BER around 3.7 mm to 4 mm for a fixed value of a fade threshold level at 1 dB. The increase in transmitter beam size decreases the wander induced pointing error. This results in decrease in scintillation index that improves the BER performance of the system up till certain optimum beam size. Further increase in beam size leads to degradation in BER performance of the system. This degradation in BER can be accounted for two reasons (i) as beam diameter becomes greater than the Fried parameter i.e., $D >> r_0$, then the beam breaks up due to loss of spatial coherence across the beam and that further increases the scintillation index, (ii) as beam diameter start approaching or becomes greater than the outer scale of turbulence i.e., $D \gtrapprox L_0$, then a small portion of the beam is diffracted and scattered independently resulting in reduction in power density and hence poor BER performance of the system. In this experiment, an increase in BER value after $W_0 = 4$ mm, clearly proves the first concept for $r_0 \sim$ 8 mm. This also indicates a degradation in Strehl ratio with increase in $D/r_0$ value as the on-axis energy would undergo random fluctuations even through beam is pointing at the distant target. For a OTG chamber dimension of 20 x 20 x 20 cm, the expected outer scale $L_0$ is $\sim$150 mm $\pm$50 mm. Therefore, as the beam diameter starts approaching outer scale, BER performance begins to degrade till it saturates when $W_0$ matches $L_0$ around 8 mm. Also, for a given transmitter beam size, the value of $r_0$ decreases with the increase in temperature that leads to increase in the value of $D/r_0$. This degrades the Strehl of the beam resulting in higher values of BER. The variation of BER with fade threshold level for various zenith angle and temperature differences is shown in Fig. 11. For a given transmitter beam size, an increase in fade threshold level decreases the BER value of the system. From Figs. 10 and 11, it is evident that BER is significantly affected by beam size, fade threshold level, and zenith angle. Also, it depends on the propagation geometry and turbulence profile. All these factors results in an optimum beam size for which the value of BER is minimum.

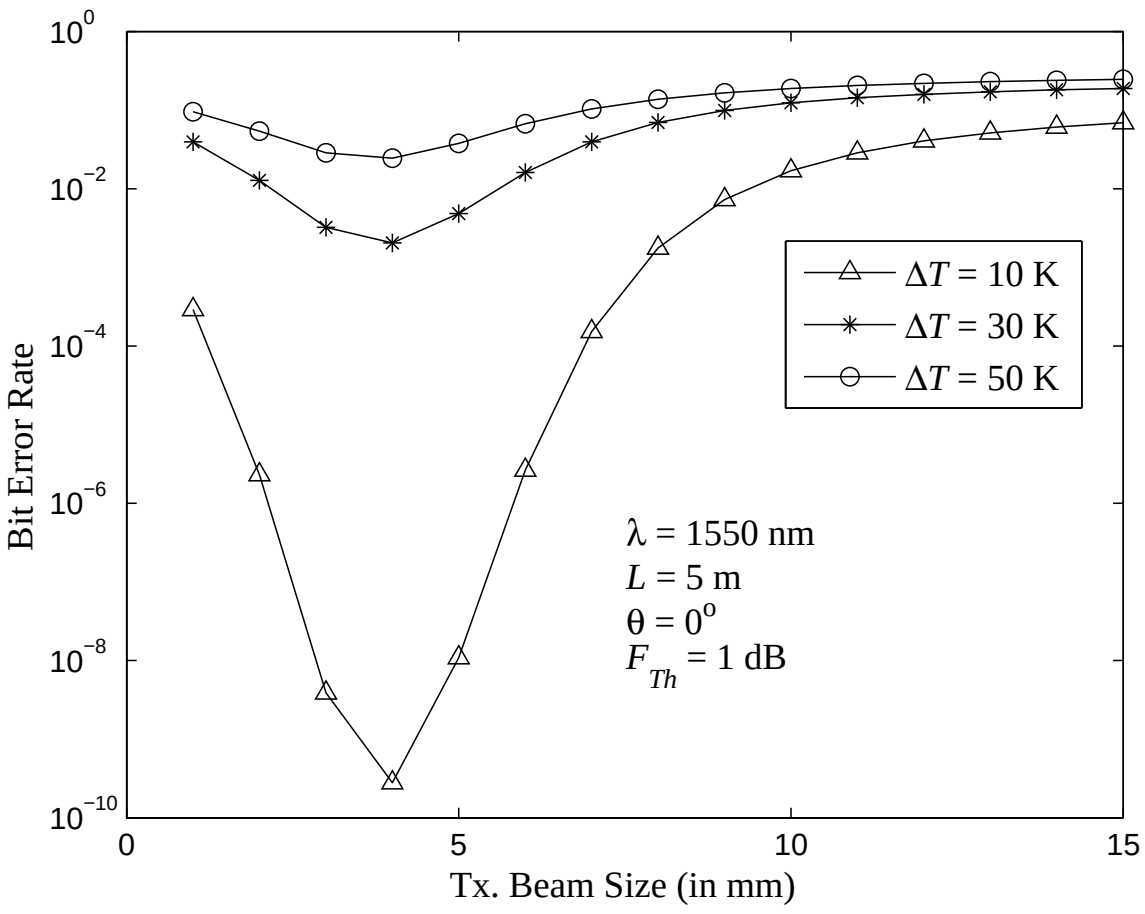

Figure 10: Variation of BER with transmitter beam size for different values of temperature difference inside OTG chamber by using semi-analytical method.

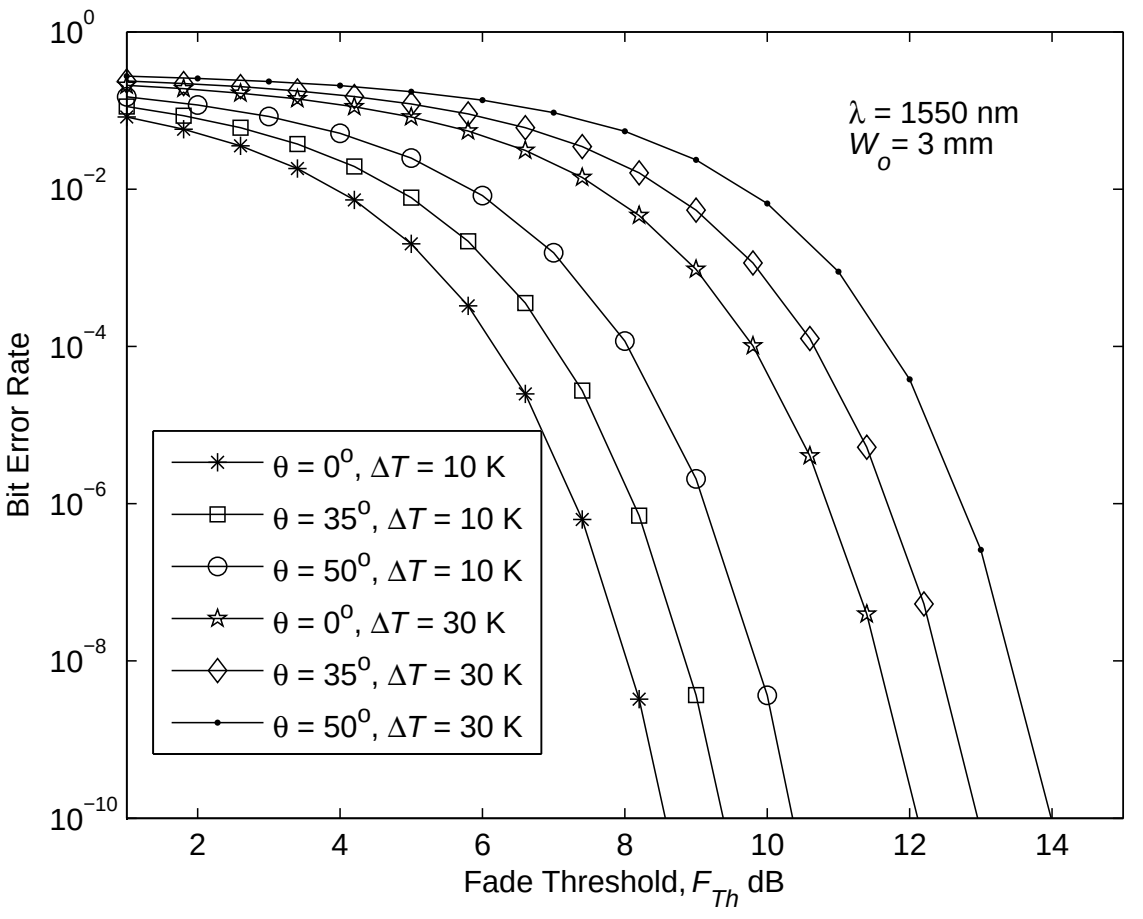

Figure 11: Variation of BER vs fade threshold level for different values of zenith angle and temperature difference inside OTG chamber by using semi-analytical method.

Figs. 12 and 13 give the variation of optimum beam size $W_{opt}$ as a function of fade threshold level $F_{Th}$ and zenith angle $\theta$, respectively for which the value of BER is minimum. It is clear from Fig. 12 that optimum beam size remains fairly constant with change in fade threshold level for low turbulence levels. However, there is a slight increase in optimum beam size for higher values of turbulence. Fig. 13 shows an increase in the value of optimum beam size with increase in zenith angle. A sharp increase in beam size is observed after zenith angle of 40 degrees.

It is clear from the results that for small propagation length $L$ and angular displacement $\alpha_r$ (i.e., $\alpha_r < \alpha_{pe}$), the radial component of scintillation index $\sigma^2_{I,r}(r, L)$ vanishes and the total scintillation index is due to longitudinal scintillation index $\sigma^2_{I,l}(L)$ only. Since longitudinal scintillation index $\sigma^2_{I,l}(L)$ is a function of induced pointing error due to beam wander, it gives a range of transmitter beam size for which scintillation index and correspondingly BER value is minimum for a given operating wavelength, fade threshold, and propagation length. It is observed that BER is a decreasing function of a fade threshold $F_{Th}$ whereas for a fixed $F_{Th}$, there exists an optimal beam size for which BER is minimum. These results are significant in the case of FSO uplink as the transmitter beam size is much smaller than the turbulent eddies near the transmitter.

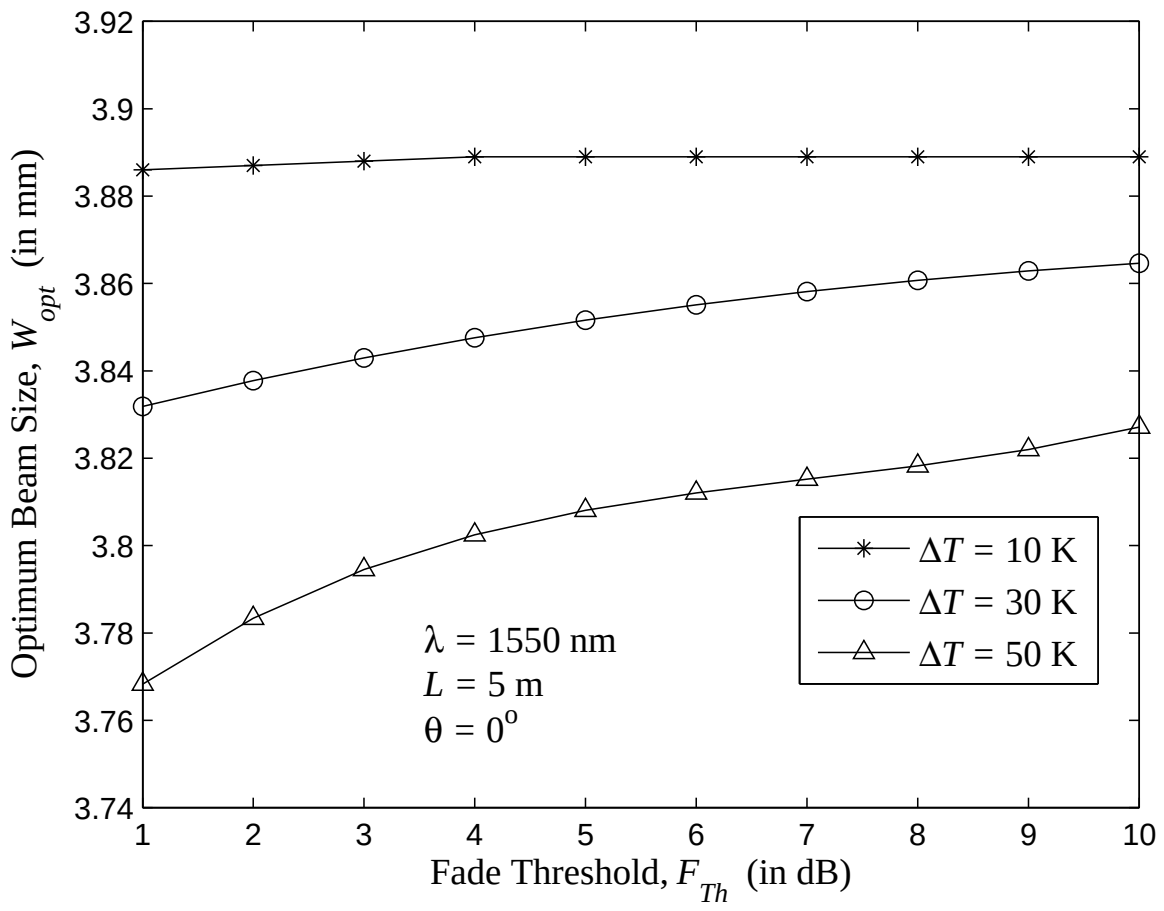

Figure 12: Variation of optimum beam size with fade threshold level for different values of temperature difference inside OTG chamber.

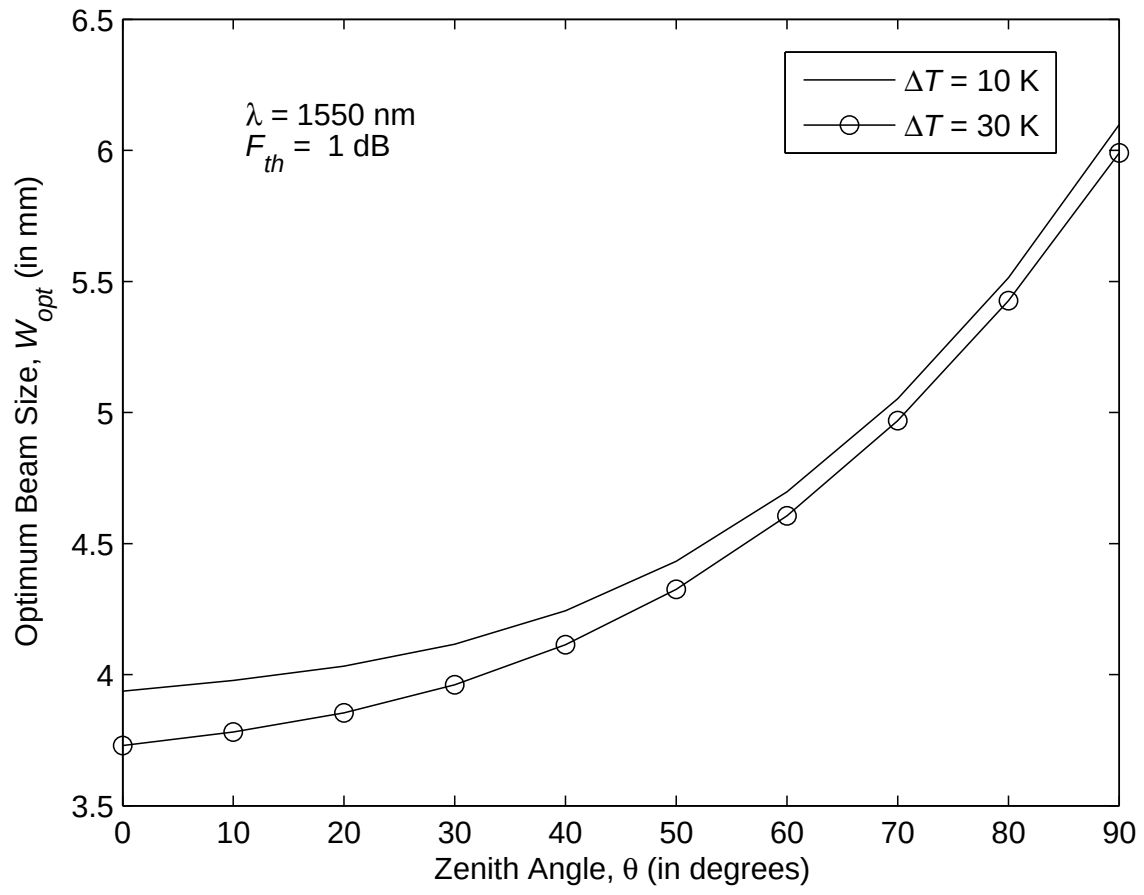

Figure 13: Variation of optimum beam size with zenith angle for different values of temperature difference inside OTG chamber.

## 5. Conclusions

In this paper, the effect of turbulence-induced beam wander variance is investigated experimentally using OTG chamber. Turbulence characterization inside OTG chamber is carried out and the results are then used to analyze the combined effect of beam wander and scintillation for FSO uplink communication. It is shown that there exists an optimum beam size for which the value of BER is minimum for a given operating wavelength, transmitter power, and link length. The optimum beam size for FSO uplink depends upon the ratio of $D/r_0$ and $D/L_0$. Increase in $D/r_0$ results in breaking up the beam into smaller pieces due to loss of spatial coherence across the beam. This degrades the Strehl of the transmitted laser beam even if the beam is pointing correctly at the target resulting in degradation in BER performance of the system. Similarly, the increase in $D/L_0$ causes reduction in the received power density due to beam diffraction and scattering, thereby increasing the value of BER till it reaches error floor at $D \approx L_0$. In order to make beam size independent of inner scale of turbulent eddies, a care must be taken to keep the ratio of $D/l_0$ at higher values so that the inner scale damping effect is almost negligible. Further, the optimum beam size changes its value with zenith angle and remains almost independent of a fade threshold level for low turbulence levels. However, a slight increase in optimum beam size is observed with a fade threshold level at higher values of atmospheric turbulence. These results will be useful while choosing system design parameters for FSO uplink communication. The work can be extended further to obtain statistical results on FSO uplink for various other experimental configurations e.g., including multiple transmitting apertures, modulation schemes or analyzing different zenith angles by increasing the propagation length within the chamber that reflects the change in propagation path in the atmosphere. This would help in better understanding of ground-to-satellite FSO uplink in varying atmospheric turbulence conditions.